\documentstyle[12pt,aasms4,psfig]{article}

\def\sun{$_\odot$}
\def\jup{$_{J}$}
\def\mum{$\mu$m}
\def\kms{km~s$^{-1}$}
\def\degree{$^\circ$}

\received{6 July 1998}
\accepted{18 August 1998}

\slugcomment{To appear in Astrophysical Journal Letters, October 1998}

\begin{document}
\title{A Candidate Protoplanet in the Taurus Star Forming Region}

\author{S. Terebey$^1$, D. Van Buren$^2$, D. L. Padgett$^2$,
T. Hancock$^1$, and M. Brundage$^2$}

\affil{$^1$Extrasolar Research Corporation, 569 S. Marengo Ave.,
    Pasadena, CA 91101}

\affil{$^2$Jet Propulsion Laboratory, IPAC 100-22, Caltech, Pasadena CA 91125}

\begin{abstract}

HST/NICMOS images of the class I protostar TMR-1 (IRAS04361+2547)
reveal a faint companion with 10.0$'' = 1400$ AU projected
separation. The central protostar is itself resolved as a close binary
with $0.31''$ = 42 AU separation, surrounded by circumstellar
reflection nebulosity. A long narrow filament seems to connect the
protobinary to the faint companion TMR-1C, suggesting a physical
association. If the sources are physically related then we hypothesize
that TMR-1C has been ejected by the protobinary. If TMR-1C has the
same age and distance as the protobinary then current models indicate
its flux is consistent with a young giant planet of several Jovian
masses.

\end{abstract}

\keywords{binaries; general --- circumstellar matter --- infrared:
stars --- planetary systems --- stars: individual (TMR-1) --- stars:
formation}

\section{Introduction}

The past few years have seen the indirect detection by their
gravitational effects of roughly one dozen extrasolar Jupiter-mass
planets around nearby stars (\cite{mar98}). Doppler surveys are
primarily sensitive to giant planets within 3 AU of the central star,
posing a challenge to theories which predict birthplaces in the 5 - 10
AU range. The conventional picture proposes a two-step formation
process where a rocky planet core forms in a disk, followed by gas
accretion over a period of 1 - 10 million years in a region outside 5
AU radius (\cite{lis95}). An alternative theory argues that giant
planets form via gravitational instabilities in the disk on a
time scale of thousands rather than
millions of years (\cite{bos98}).

We have detected a low-luminosity object near the class I protostar 
TMR-1 whose flux is consistent with that of a giant protoplanet. If 
confirmed, the protostar's age of approximately 300,000 years places 
severe constraints on the time scale of giant planet formation. 
Furthermore the data show that the TMR-1 protostar is a binary system. 
Most stars are members of multiple star systems with separations ranging 
widely from sub-AU to thousands of AU; the overlap with circumstellar disk
sizes has consequences for planet formation (\cite{ben96}; \cite{bat97}).  
An important issue is whether binary stars provide a hospitable environment
for the formation of substellar mass objects.

\section{TMR-1 protostars}

Previous observations establish TMR-1 as a typical class I protostar, 
similar in mass and luminosity to the Sun having $\sim 0.5$ M\sun\ 
and 3.8 L\sun, respectively, and which millimeter observations 
suggest has a low-mass disk (\cite{bon96}). Observed NIR magnitudes 
are J~=~16.1, H~=~12.9, and K~=~10.6 (\cite{ter90}; \cite{ken93}). 
Class I protostars are surrounded by opaque envelopes of infalling 
gas and dust. Based on statistical and theoretical arguments typical 
ages of class I protostars are 100,000 to 300,000 years ({\cite{ter84};
{\cite{shu87}). NIR imaging and millimeter interferometry data of
TMR-1 show a bipolar outflow which extends SE to NW (\cite{ter90};
\cite{hog98}). Based on the arguments given in Chandler et al.\ (1996)
TMR-1 is not viewed edge-on or pole-on, but at an intermediate ($\sim
60$\degree) inclination.

\section{HST NICMOS Images and Photometry}

The high spatial resolution ($0.15''$ at 1.6 \mum) HST/NICMOS images
in Figure 1 resolve the TMR-1 protostar into two point sources which
we call A and B.  A is the northern component.  At the Taurus cloud
distance of 140 pc the $0.31''$ projected separation is 42 AU, a
fairly typical binary separation. The new data reveal TMR-1 to be a
protobinary surrounded by gas and dust, viewed during the epoch of
formation.

Figure 1 displays extensive nebulosity, brightest near the
protobinary. A long narrow filament extends in a gentle curve from
near the protobinary to a third fainter point source, which we call 
C, located $10.0''$ southeast. The image provides strong visual 
evidence that object C appears associated with the
protobinary by means of the filament. TMR-1C is detected at S/N = 50
in the F205W filter, as implied by the presence of the Airy
diffraction ring.

The image artefacts (\cite{cas97}) are easily identified in the
original image orientation (+y axis at P.A. = 38\degree\ E of
N). Artefacts arising from the bright protostars include: the $\pm
45$\degree telescope diffraction spikes; electronic ghost stars at
$\pm 128$ pixels along the x,y axes; and two faint electronic ghost
columns, one of which passes through the protostars, and  another
seen 128 pixels to the left. Finally, a residual coronographic spot 
appears in the upper left hand quadrant.

Extended nebulosity is common around protostars at near-infrared
wavelengths. Stellar photons escape through the transparent polar
regions created by the bipolar outflow; they delimit the $\tau \sim 1$
surface when they absorb or scatter in the dusty infall envelope and
dusty circumstellar disk (\cite{whi97}). The highly structured
nebulosity around the TMR-1 protostars implies the density is
inhomogeneous. The sharp contrast of the filament above the
background suggests the density is locally enhanced, while the
illumination of the filament at large (10$''$) distances suggests a
fairly clear line of sight back to the protostars.

Table \ref{tbl-1} gives positions and individual component fluxes via
PSF fitting. Fluxes based on aperture photometry are given in Table
\ref{tbl-2} both to facilitate comparison with ground based
measurements and to place the data on the STScI HST/NICMOS photometric
system. Ground-based 2.2\mum\ K-band IRTF images and also K$'$ at Keck
(\cite{bla98};\cite{hog98}) confirm the detection and approximate flux of object C.

\section{Local NIR Star Counts, Chance Background Object, and Extinction \label{sec-bkg}}

For the Taurus cloud K-band star counts give $N(K' < K) = 0.041 \times
10^{0.32K}$ stars per square degree, which includes an extinction of
A$_K = 0.4$ estimated from the same data (\cite{bei94}). Assuming 18.5
for the K-band magnitude implies about one background star per NICMOS
frame.

{\it A posteriori} probability estimates are problematic. However we press
on noting that the TMR-1 filament is a unique structure in our
ensemble of HST/NICMOS images. The chance that a random background
star lies at the tip of the filament is 2\% if we assign a
conservatively large $3'' \times 3''$ effective search area.

The scarcity of background stars is empirically confirmed by the HST
data, which show fewer than expected background stars because of high
extinction local to the protostars. Comparable S/N HST/NICMOS images
for nine class I protostars in Taurus show one other secondary object 
(K = 18.7 mag), giving one or two possible background objects in nine
fields. To match the large scale NIR star counts implies an average
extinction over the 20$''$ NICMOS field of view of A$_K = 1$ to 2
(A$_V = 10$ to 20) towards Class I protostars in Taurus.

An alternate estimate for the extinction is set by values previously
derived for the protostar, which range from 2.5 to 4 at K
(\cite{ter90}; \cite{whi97}). The extinction is likely smaller at
10$''$ distance from the protostar, as is also suggested by the Table
\ref{tbl-1} flux ratios. Intrinsic NIR stellar colors are near zero
because the spectral energy distribution of many stars is near the
Rayleigh-Jeans limit. The observed highly reddened colors of
protostars are therefore caused by extinction and scattering. The
increasing flux ratio of object C to either protostar between 1.6 and 
2.05 \mum\ suggests less extinction toward object C than toward A and B.

\section{Luminosity and Temperature}

Models of giant planet and brown dwarfs imply they are hottest and
brightest when young, as luminous as 0.01 L\sun\ at one million years
(\cite{nel93}; \cite{bur97}). The radii are near that of Jupiter's,
R\jup\ $ = 7.1 \times 10^9$ cm, over a large mass and
temperature range; young objects should be modestly (up to factor of three)
larger. Models suggest effective temperatures as great as 3000 K below
one million years age.

The object TMR-1C is clearly much fainter than the neighboring
protostars; if located at the same distance as the Taurus cloud then
the estimated bolometric luminosity is approximately $10^{-3}$ to
$10^{-4}$ L\sun, within the giant planet to brown dwarf regime. To
derive its luminosity the observed NIR fluxes were fit assuming for
simplicity a black-body spectrum extincted by dust. Dust extinction
parameters are from Draine (1998). Assumptions are 140 pc distance, 1
R\jup\ minimum radius for stellar and substellar objects, and A$_V <
30$ (A$_K < 3$) extinction. The extinction cap is selected as the
maximum compatible with NIR background source counts (Section
\ref{sec-bkg}).

General results from varying T$_{eff}$, A$_V$, and radius are that the
temperature is not well constrained as values ranging from 1200 K
(A$_V = 0$) to 3000 K (A$_V = 30$) give acceptable fits. However the
radius is reasonably constrained to be a few R\jup\ at the assumed
distance, and depending also on the maximum extinction at the high
temperature end. Relaxing our assumptions, hotter background stars
provide acceptable fits if higher extinctions are allowed. Foreground
stars or low extinctions are ruled out; the approximate H - K color of
1.5 is redder than the photospheres of known low luminosity stars.

The broadband NICMOS filters give limited spectral information but
allow us to exclude effective temperatures below approximately 1600 K
since there is no evidence in TMR-1C for a strong methane dip near 1.8
microns (e.g. \cite{all95}). Near-infrared spectra of cool objects
($\sim 2000$ K) which show water at 1.9 \mum\ are sufficiently
featureless to be consistent with our photometry (e.g. Figure 7 of
\cite{opp98}). Better constraints on the extinction and effective
temperature await low resolution spectra of TMR-1C.

\section{Mass}

Model evolutionary tracks for giant planets and brown dwarfs show the
derived mass depends strongly on the age and luminosity (Figure 7 of
\cite{bur97}). If TMR-1C has the same 300,000 year age assumed for the
protostars then A$_V$ is 8 - 20 and the mass is 2 - 5 M$_J$. If the
age is ten million years, the same as older pre-main sequence stars in
Taurus, the mass may be as high as 15 M$_J$. However, below one
million years the models are sensitive to the initial conditions, as
the thermal relaxation timescale is comparable to the planet's
age. More realistic models will depend on the planetary formation
mechanism.

\section{Ejection Hypothesis}

If TMR-1C is a physical companion of the TMR-1 binary then models suggest 
it formed much closer to the protostars than its observed 1400 AU 
projected distance. We hypothesize TMR-1C has been ejected by the two
protostars.  Apart from some exceptions such as hierarchical systems,
celestial dynamics finds that 3-body stellar systems with comparable
separations are unstable and tend to eject the lowest mass object
(\cite{mon76}). On dimensional grounds the characteristic velocity of
ejection is $(G M / R)^{0.5}(1+e)$, the velocity of periastron passage
of the binary. Numerical studies show a large dispersion in ejection
velocities (\cite{sta72}).

The separation of the protostars allows us to estimate a
characteristic ejection velocity. The computation is only indicative
given that the orbital parameters and inclination are poorly
known. The observed projected separation of stars A and B is 42 AU;
statistically binaries spend most time at the widest separations.  For
a typical binary eccentricity of $e = 0.5$ the separation varies by a
factor of three. Including a modest deprojection correction,
periastron passage may occur at 15 - 30 AU separation.  The
corresponding ejection velocity is 5 - 10 \kms\ for 1 M\sun\ assumed
total mass. The current distance of $10''$ then implies the time since
ejection is about 1000 yr.

Consider for the moment that the filament marks the trail of object C.
The filament's shape is curved and appears consistent with the
expected hyperbolic trajectory. However shear is likely important if
the filament lies within the differentially rotating infall envelope
or disk. The assumption of Keplerian rotation is adequate to estimate
the timescale (\cite{ter84}). The period is 1000 yr at 100 AU radius,
which implies significant wrapping can be expected on roughly $0.67''$
size scales.

\section{Filament}

Although the position angles of the filament and outflow are similar,
the filament differs from typical NIR outflow structures. Models of 
outflow cavities show conical shapes (\cite{whi97}); if the outflow 
cavity is limb-brightened it should have two symmetric horns with a 
sharp outer edge, whereas what is observed is one filament whose sharp
edge is on the wrong (southern) side given its curvature. NIR
polarimetry data show the filament is primarily scattered stellar
light emanating from the protostars (\cite{luc97}) which rules out an
emission line jet.

The filament is projected against the outflow but the moderate source
inclination implies the filament could traverse either the outflow
cavity or the dense infall envelope. One possible explanation is the 
filament may be a material tail, such as for example a tidal tail 
formed by two colliding circumstellar disks ({\cite{lin98}). Ground-based 
data in support of a material tail show HCO$^+$ along the filament, 
indicating the presence of dense gas (\cite{hog98}). However HCO$^+$ 
can be ambiguous as a dense gas tracer since it often has enhanced 
abundance in molecular outflows. Alternatively the filament may be an 
illumination channel, or light pipe, created when the protoplanet 
tunneled through the infall envelope. A drawback to the light-pipe 
explanation is that Bondi-Hoyle gravitational accretion implies a 
diameter which is too narrow to explain the observed filament so some 
other mechanism must operate.

\section{Isolated Planets}

We have proposed that TMR-1C is a substellar object which has been
ejected by a binary protostar. There are two key experiments to test
the idea that TMR-1C is an ejected protoplanet. Spectra will measure
the extinction and effective temperature to better discriminate
between stellar, brown dwarf, or planet origin. In several years proper 
motion measurements will detect TMR-1C's motion on the sky.  The 
predicted direction may be along, or in the case of a tidal tail, at 
an angle to the filament ({\cite{lin98}).

We outline one of the many possible mechanisms for planet
ejection. Three-body numerical simulations suggest stable planetary
orbits exist at radii approaching half the binary periastron
separation (\cite{ben96}). In other words there is a maximum stable
radius for planet formation in a binary system. A substellar object that migrates or forms in the zone of marginal stability is subject to orbital
resonance pumping. After repeated periastron passages the object gains
sufficient energy to escape the system. This mechanism does not
require a gaseous disk per se, and so may apply to pre-main sequence
stars as well as protostars.

The discovery of an ejected protoplanet is unexpected. However, given
the prevalence of binary systems the process seems inevitable, and the
question becomes how often. The idea that young planets should
occasionally be ejected from their solar systems is rich in
implications, both for our understanding of how planetary systems
form, and in strategies for detecting isolated planets using
current technology.

\acknowledgments Many people provided support or encouragement.  A
special thanks to Charlie Lada for pointing out that brown dwarfs
would have K magnitude near 17 in Taurus. We thank John Rayner and Bob
Joseph for providing Infrared Telescope Facility observations on short
notice. Terebey gratefully acknowledges NASA support including NASA
Origins of Solar Systems Program funding under contract NASW-97009,
and funding from grant GO-07325.01-96A through the Space Telescope
Science Institute, which is operated by the Association of
Universities for Research in Astronomy, Inc. under NASA contract
NAS5-26555. This work was carried out in part at the Jet Propulsion
Laboratory, operated by the California Institute of Technology under
contract for NASA.

\clearpage

\figcaption{True-color NICMOS HST image of protostar TMR-1 with $3'' = 420$ AU scale.   There is a long narrow filament extending from the protobinary to a faint point source which is the candidate protoplanet. The inset shows a magnified view ($0.3''$ scale) of the protobinary in the central region, complete with first Airy diffraction ring around each component. The field is suffused by bright nebulosity. The region is heavily reddened, but short wavelengths have been boosted to make the nebulosity appear white. Red (2.05
\mum = F205W filter); green (1.87 \mum = F187W filter); blue (1.60 \mum = 
F160W filter); North is up. Display uses log stretch. 
\label{fig1}}

\clearpage

\begin{deluxetable}{crrrrrrrrrrr}
\footnotesize
\tablecaption{TMR-1 HST NICMOS Positions$^a$ and Photometry$^b$. \label{tbl-1}}
\tablewidth{0pt}
\tablehead{
\colhead{Component} & \colhead{Right Ascension} & \colhead{Declination} &\colhead{F160W}  &\colhead{F187W} &\colhead{F205W}  \nl
&\colhead{(J2000)} & \colhead{(J2000)} & \colhead{1.60 \mum}  & \colhead{1.87 \mum} & \colhead{2.05 \mum} \nl
}
\startdata
A & 04h39m13.84s & $25^{\circ}53'20.6''$        &$38^{+22}_{-20}$               &$53^{+10}_{-11}$               &$250^{+80}_{-70}$      \nl
B & 04h39m13.83s & $25^{\circ}53'20.4''$        &$9^{+17}_{-9}$                 &$23.5^{+8}_{-8}$               &$180^{+70}_{-117}$             \nl
C & 04h39m14.14s & $25^{\circ}53'11.8''$        &$1.17^{+0.06}_{-0.04}$ &$0.98^{+0.05}_{-0.04}$ &$3.32^{+0.20}_{-0.20}$ \nl
\enddata


\tablenotetext{a}{$0.35'' = 1 \sigma$ HST absolute position error.}
\tablenotetext{b}{PSF amplitude fitting and $1 \sigma$ limits in Counts/s image units. Limits are instrumental for C, dominated by background structure for A and B. Absolute photometry errors of 3\% should be added in quadrature. Applying
no color correction, to convert units to $\mu$Jy multiply by 15.6, 34.4, and 12.4 for F160W, F187W, and F205W filters, respectively.}

\end{deluxetable}

\clearpage

\begin{deluxetable}{crrrrrrrrrrr}
\footnotesize
\tablecaption{Flux of TMR-1C$^a$. \label{tbl-2}}
\tablewidth{0pt}
\tablehead{
\colhead{R} &  \colhead{J}      & \colhead{F160W} &
\colhead{F187W}  &  \colhead{F205W} & \colhead{K} & \colhead{L} & \colhead{Units}                 \nl
}
\startdata
 .. &  ..   & 16. & 33. & 39. & .. & .. & $\mu$Jy \nl
 $>22.2^{b}$ & $>21.2^{c}$ & 19.6 & 18.5 & 18.2 & $17.9^{c}$ & $>13.2^{c}$ &mag  \nl
\enddata

\tablenotetext{a} {NICMOS magnitudes are based on standard $0.5''$ apertures and the HST Vega system. Background structure dominates the 10\% photometric uncertainty.}
\tablenotetext{b}{ $5\sigma$\ limit from \cite{jar94}.}
\tablenotetext{c}{April 1998 IRTF NSFCAM data. K band has 20\% photometric uncertainty; $3\sigma$\ limits elsewhere.}

\end{deluxetable}

\clearpage

\psfig{file=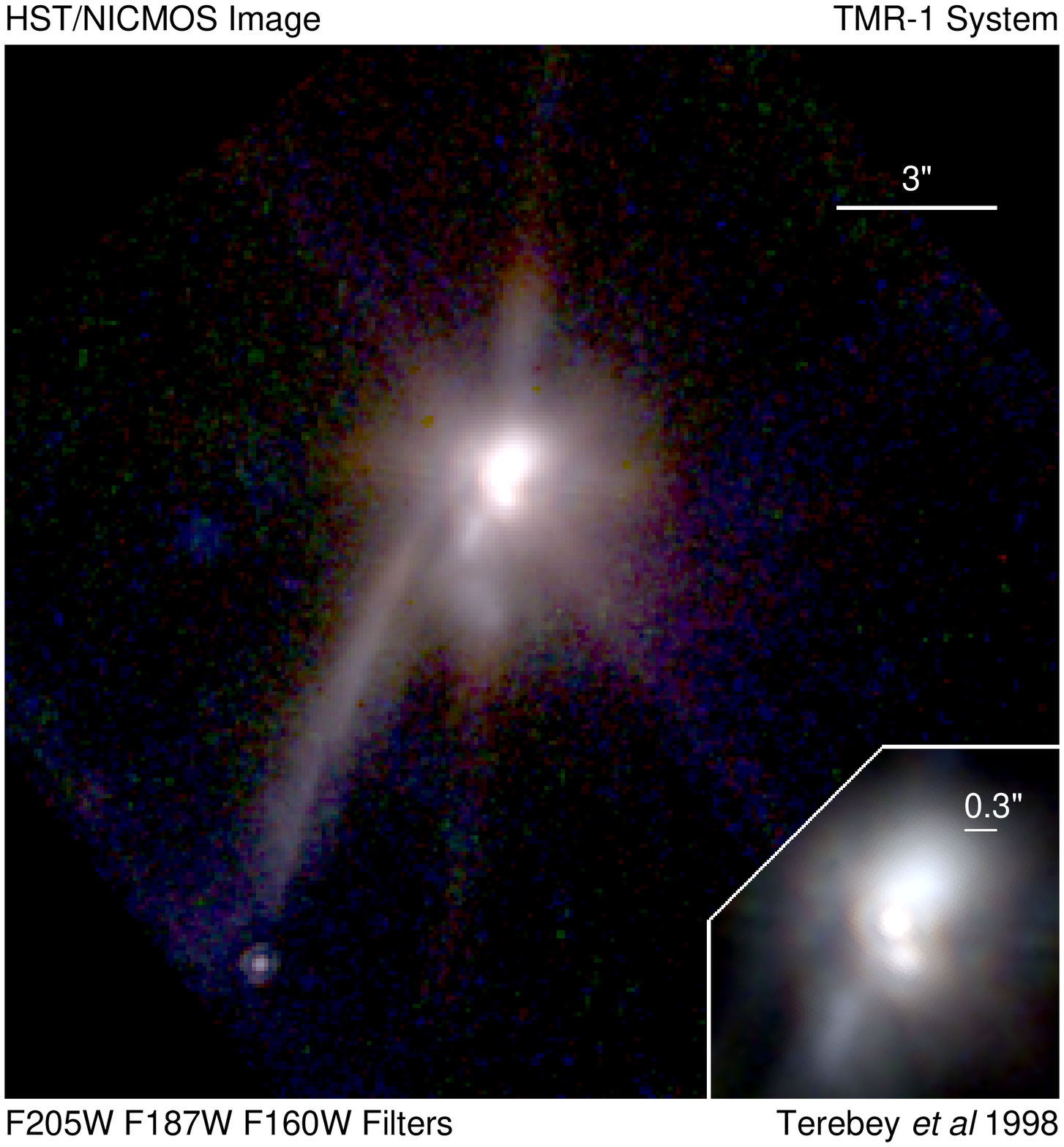}


\begin{thebibliography}{}

\bibitem[Allard \& Hauschildt 1995] {all95} Allard, F. \& Hauschildt, P. H. 1995, ApJ, 445, 433
\bibitem[Bate \& Bonnell 1997] {bat97} Bate, M. R. \& Bonnell, I. A. 1997, MNRAS, 285, 33
\bibitem[Beichman \& Jarrett 1994] {bei94} Beichman, C. A., \& Jarrett, T. 1994, Astro. \& Spa. Sc., 217, 207
\bibitem[Benest 1996] {ben96} Benest, D. 1996, A\&A, 314, 983
\bibitem[Blake 1998] {bla98} Blake, G. 1998, private communication
\bibitem[Bontemps et al.\ 1996] {bon96} Bontemps, S., Andr\'e, P., Terebey, S., \& Cabrit, S. 1996, A\&A, 858
\bibitem[Boss 1998]{bos98} Boss, A. P. 1998, Nature, 393, 141
\bibitem[Burrows et al.\ 1997] {bur97} Burrows, A., Marley, M., Hubbard, W.B., Lunine, J. I., Guillot, T., Saumon, D., Freedman, R., Sudarsky, D., \& Sharp, C. 1997, ApJ, 491, 856
\bibitem[Chandler et al.\ 1996] {cha96} Chandler, C. J., Terebey, S., Barsony, M., Moore, T. J. T., \& Gautier, T. N. 1996, ApJ, 471, 308
\bibitem[Casertano 1997] {cas97} The 1997 HST Calibration Workshop With a New Generation of Instruments'', proceedings of a meeting held Sept 22 - 24, 1997 at the Space Telescope Science Institute in Baltimore, MD. Eds. S. Casertano, R. Jedrzejewski, T. Keyes, \& M. Stevens, (Space Telescope Science Institute: Baltimore), Available on-line at http://www.stsci.edu
\bibitem[Draine 1998] {dra98} Draine, B. 1998, private communication
\bibitem[Hogerheijde et al.\ 1998] {hog98} Hogerheijde, M. R., van Dishoeck, E. F., Blake, G., \& van Langevelde, H. J. 1998, ApJ in press.
\bibitem[Jarrett et al.\ 1994]{jar94} Jarrett, T. H., Dickman, R. L. \& Herbst, W. 1994, ApJ, 424, 852
\bibitem[Kenyon et al.\ 1993] {ken93} Kenyon, S. J., Whitney, B. A., Gomez, M., \& Hartmann, L. 1993, ApJ, 414,773
\bibitem[Lin et al\ 1998] {lin98} Lin, D. N. C., Laughlin, G., Bodenheimer, P., \& R\'ozyczka, M. 1998, Science, in press
\bibitem[Lissauer 1995] {lis95} Lissauer, J. J. 1995, Icarus, 114, 217
\bibitem[Lucas \& Roche 1997] {luc97} Lucas, P. W. \& Roche, P. F. 1997, MNRAS, 286, 895
\bibitem[Marcy \& Butler 1998]{mar98} Marcy, G. W. \& Butler, R. P. 1998, ARAA, 36, 57
\bibitem[Monaghan 1976]{mon76} Monaghan, J. J. 1976, MNRAS, 176, 63
\bibitem[Nelson, Rappaport, \& Joss 1993] {nel93} Nelson, J. A., Rappaport,S., \& Joss, P. C. 1993, ApJ, 404, 723
\bibitem[Oppenheimer et al.\ 1998] {opp98} Oppenheimer, B. R., Kulkarni, S. R., Matthews, K., \& van Kerkwijk, M. H. 1998, ApJ, in press
\bibitem[Shu, Adams, \& Lizano 1987] {shu87} Shu, F. H., Adams, F. \& Lizano, S. 1987, ARA\&A, 25, 23
\bibitem[Standish 1972] {sta72} Standish, E. M. 1972, A\&A, 21,185
\bibitem[Terebey et al.\ 1990] {ter90} Terebey, S., Beichman, C. A., Gautier, T. N., \& Hester, J. J. 1990, ApJ, 362, L6
\bibitem[Terebey et al.\ 1984] {ter84} Terebey, S., Shu, F. H., \& Cassen, P. 1984, ApJ, 286, 529
\bibitem[Whitney et al.\ 1997] {whi97} Whitney, B. A., Kenyon, S. J., \& Gomez, M. 1997, ApJ, 485, 703

\end{thebibliography}
\end{document}